\documentstyle[prb,aps,graphicx]{revtex}

\begin{document}
\draft 
\twocolumn [\hsize\textwidth\columnwidth\hsize\csname
@twocolumnfalse\endcsname

\title{Temperature dependence of single particle excitations in a ${\bf
\textit{S}=1}$ chain: exact diagonalization calculations compared
to neutron scattering experiments}

\author{M. Kenzelmann and P. Santini}

\address{Oxford Physics, Clarendon Laboratory, Oxford OX1 3PU, United Kingdom}
\date{\today}
\maketitle

\begin{abstract}
Exact diagonalization calculations of finite antiferromagnetic
spin-1 Heisenberg chains at finite temperatures are presented and
compared to a recent inelastic neutron scattering experiment for
temperatures $T$ up to $7.5$ times the intrachain exchange
constant $J$. The calculations show that the excitations at the
antiferromagnetic point $q=1$ and at $q=0.5$ remain resonant up to
at least $T=2J$, confirming the recent experimental observation of
resonant high-temperature domain wall excitations. The predicted
first and second moments are in good agreement with experiment,
except at temperatures where three-dimensional spin correlations
are most important. The ratio of the structure factors at $q=1$
and at $q=0.5$ is well predicted for the paramagnetic
infinite-temperature limit. For $T\leq2J$, however, we found that
the experimentally observed intensity is considerably less than
predicted. This suggests that domain wall excitations on different
chains interact up to temperatures of the order of the spin band
width.
\end{abstract}

\pacs{PACS numbers: 75.10.Jm, 75.40.Gb, 75.40.Mg} ]

\section{Introduction}
The properties of antiferromagnetic spin-1 Heisenberg chains
result from the presence of strong quantum fluctuations in the
ground state, preventing magnetic order even at zero temperature
where thermal fluctuations are absent. The ground state is a
macroscopic quantum state, a non-magnetic spin singlet, where the
spin correlation function falls off exponentially with increasing
distance.\cite{Haldane83} At zero temperature an antiferromagnetic
spin-1 chain has a hidden long-ranged spin
order\cite{Den_Nijs_Rommelse,White_Huse} which can carry
long-lived spin-1 triplet excitations despite the absence of
conventional magnetic long-range order. These excitations can be
regarded as domain walls\cite{Fath_Solyom} of the hidden spin
order and they have a mass $\Delta = 0.41 J$ ($J$ is the exchange
constant) which is generated by strong quantum fluctuations and is
not a result of an anisotropy in the
system.\cite{Haldane83,Buyers86,Nightingale_Blote} The excitation
spectrum of an antiferromagnetic spin-1 Heisenberg chain is
fundamentally different from the equivalent spin-1/2 chain whose
excitations are gapless spin-1/2 particles.\cite{Tennant93}\par

At finite temperature, temperature fluctuations disorder the
hidden "quantum order" and this leads to a renormalization of the
triplet excitations. For $T<\Delta$, the triplet excitations are
well-defined and the spin-wave velocity is constant, but their
mass - defined as the excitation energy at the antiferromagnetic
point - increases.\cite{Kenzelmann_CsNiCl3_gap} This upward
renormalization of the excitation energy is well described by a
low-energy field theory, the non-linear sigma
model.\cite{Jolicoeur_Golinelli,Kenzelmann_CsNiCl3_gap,Kenzelmann_CsNiCl3_temperature}
For finite temperatures, the excitations acquire a finite
life-time due to collisions with thermally excited particles and
the temperature dependence of the decreasing life-time of the
excitations is well described by a semi-classical model of gapped
spin-1 particles with a relativistic
dispersion.\cite{Damle_Sachdev,Kenzelmann_CsNiCl3_gap,Kenzelmann_CsNiCl3_temperature}\par

For $T>\Delta$, there is a cross-over to a qualitatively different
behavior. In this temperature range the excitations are resonant
at least up to $T=2.7J$,\cite{Kenzelmann_CsNiCl3_gap} but the
excitation width is nearly as large as the excitation energy and
the spin-wave velocity decreases with increasing temperature.
These high-temperature excitations can be regarded as domain walls
of a local hidden spin
order\cite{Yamamoto_Miyashita,Kenzelmann_CsNiCl3_temperature}
which is the remainder of the hidden long-range order at
$T=0\;\mathrm{K}$. However, a quantitative theoretical
understanding of the high-temperature spin dynamics is missing at
present.\par

In the lack of reliable analytical methods for a quantitative
calculation of dynamical properties at finite $T>\Delta$ we
calculated the scattering function numerically by exact
diagonalization (ED). We considered finite spin-1 chains with an
antiferromagnetic Heisenberg exchange, described by the
Hamiltonian
\begin{equation}
    {\mathcal{H}}=J \sum_{i} \bbox{S}_{\rm i} \cdot
    \bbox{S}_{\rm i+1}\, ,
    \label{Hamiltonian1D-diag}
\end{equation}and calculated the dynamic structure factor and
frequency moments for temperatures up to $T=7.5J$. The results of
the calculations were then compared with the measured neutron
scattering spectrum of the antiferromagnetic spin-1 Heisenberg
chain compound ${\rm CsNiCl_3}$.\par

${\rm CsNiCl_{3}}$ crystallizes in a hexagonal crystal structure,
$D^{4}_{6h}$ space group and is one of the best model systems for
investigating antiferromagnetic spin-1 Heisenberg chains. The
spin-1 components are carried by ${\rm Ni^{+2}}$ which interact
via a super-exchange involving ${\rm Cl^{-}}$-ions. The
super-exchange interaction $J$ along the c-axis is much stronger
than the super-exchange interaction $J'$ in the basal plane. Thus
${\rm CsNiCl_{3}}$ approximates a system of weakly coupled spin-1
chains. The Hamiltonian can be written as
\begin{eqnarray}
    &{\mathcal{H}}_{\rm {\tiny CsNiCl_{3}}}=J
    {\displaystyle \sum_{i}} \bbox{S}_{i} \cdot
    \bbox{S}_{i+1}&  \nonumber \\ & + J'
    {\displaystyle \sum_{<i,j>}} \bbox{S}_{i}
    \cdot \bbox{S}_{j} - D \sum_{i} (\bbox{S_{i}^{z}})^2\, ,&
    \label{Hamiltonian-diag}
\end{eqnarray}with $J=2.28\;\mathrm{meV}$ and
$J'=0.044\;\mathrm{meV}$.\cite{Katori,Buyers86,Morra} The sum in
the basal plane includes the interaction between nearest-neighbors
$<i,j>$ only. The single-ion crystal-field anisotropy
$D=4\;\mathrm{\mu eV}$ is small enough that the exchange
interaction between the spins is isotropic in spin space. A
detailed description of the neutron experiments has been presented
elsewhere.\cite{Kenzelmann_CsNiCl3_gap,Kenzelmann_CsNiCl3_temperature,Kenzelmann_CsNiCl3_continuum,Kenzelmann_CsNiCl3_continuum_long}\par

This paper is structured as follows: Section II introduces the
numerical calculations, and shows that the excitations are
resonant up to at least $T=2J$. In Section III energy-integrated
quantities of the dynamic structure factor are quantitatively
compared with the experiment and effects of biquadratic and
next-nearest neighbor interactions on the excitations are
discussed. Section IV addresses the temperature dependence of the
expectation value of the Hamiltonian and shows that there is
excellent agreement between experiment and theory. Section V
summarizes the results.\par

\section{Exact diagonalization}
The scattering function was obtained numerically by ED for single
spin-1 chains of finite length through the following formulae:
\begin{equation}
S^{zz}(q,\omega) = \frac{1}{\pi} \frac{1}{1-\exp{(-\beta
\omega})}\Im(\chi^{zz}(q,\omega)) \nonumber
\end{equation}
\begin{eqnarray} &\Im(\chi^{zz}(q,\omega))=& \nonumber \\ \nonumber \\&\pi
{\displaystyle \sum_{i,j}}\vert\langle i|S_z(q)|j\rangle\vert^2
(n_i - n_j)\delta(\omega-(\omega_j - \omega_i))
\label{Eq_matrix_elements}
\end{eqnarray}\begin{equation}
S_z(q)=\frac{1}{\sqrt{N_s}}\sum_R S_z(R) e^{i q R}, \nonumber
\end{equation}where $|i\rangle$ are the exact eigenstates of the
Hamiltonian, with energy $\omega_i$, and Boltzmann population
factors $n_i$. $N_s$ is the number of spins in the chain, and
$S_z(R)$ is the operator for the $z$ component of the spin at
position $R$. $q$ is the wave-vector along the chain.\par

The calculation of finite temperature properties requires
knowledge of the full spectrum of the Hamiltonian so that only
rather small sizes can be handled. We used a chain with $N_s = 8$
sites with periodic boundary conditions. The dimension of the
Hilbert space is $3^{N_s} = 6561$. To reduce the numerical effort,
we exploit the conservation of the $z$ component of the total spin
$S_z^{\rm tot}=\sum_R S_z(R)$, which makes the Hamiltonian matrix
block-diagonal in the basis of eigenstates of $S_z^{\rm tot}$.
Since the operator $S_z(q)$ commutes with $S_z^{\rm tot}$, matrix
elements in Eq.~\ref{Eq_matrix_elements} are nonzero only if $|i>$
and $|j>$ have the same value of $S_z^{\rm tot}$. Thus, once the
partition function (entering $n_i$ and $n_j$) has been determined,
the contribution of each $S_z^{\rm tot}$ eigenspace to the
scattering function can be calculated independently.\par

\begin{figure}
\begin{center}
  \includegraphics[height=10cm,bbllx=80,bblly=90,bburx=530,
  bbury=625,angle=0,clip=]{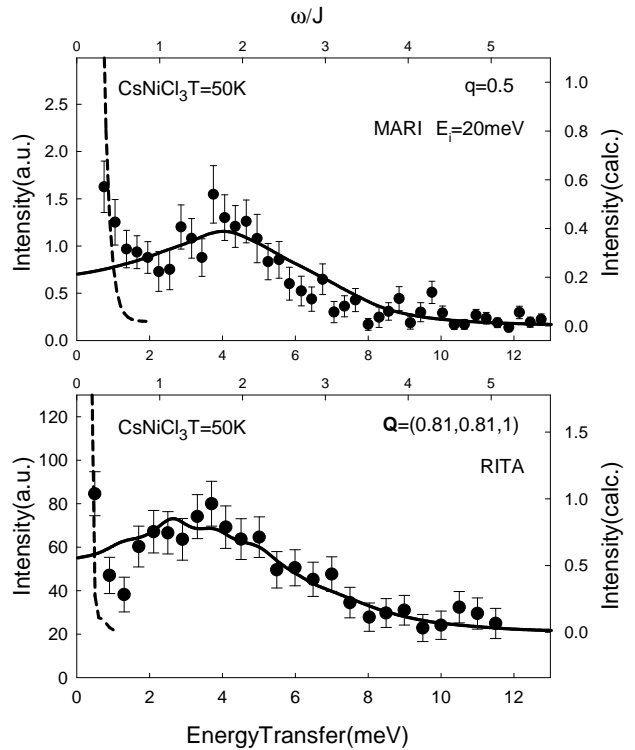}
  \vspace{0.3cm}
  \caption{The solid line represents the calculated dynamic
  structure factor $S^{zz}(q,\omega)$ for $q=0.5$ (upper figure) and
  $1$ (lower figure) at $T=2J$ as a function of energy $\omega/J$
  (top axes). The calculations for a finite chain of $8$ sites
  were convoluted with a Lorentzian with a half width $0.3 J$ in
  order to smooth out finite-size oscillations. The intensity is
  for a chain of $N_s=8$ spins and such that
  $\sum_{k_i} \int d\omega S^{zz}(k_i,\omega)=\frac{S(S+1)N_s}{3}$ for
  $S=1$ (right axes). The measured neutron scattering
  spectra are shown as solid circles and as a function of neutron
  energy transfer $\omega$ (bottom axes). The data were measured
  at $50\;\mathrm{K} \simeq 2J$ using the RITA and the MARI
  spectrometer, taken from Refs.~\cite{Kenzelmann_CsNiCl3_gap,Kenzelmann_CsNiCl3_temperature},
  and in units shown on the left axes. The axes on the left
  and right in the two figures were offset to account for the
  experimental background scattering. The dashed lines
  corresponds to the elastic scattering peak (not completely
  shown) which arises due to incoherent scattering at the sample
   and its environment.}
  \label{Fig_Paolo_spectrum}
\end{center}
\end{figure}

Zero-temperature results have been obtained for sizes up to
$N_s=18$ by using the Lanczos algorithm to find the exact ground
state of the system. In these calculations, the Hamiltonian matrix
is further reduced to block-diagonal form by exploiting
translational invariance, using a basis of states with given total
wave-vector $K_{\rm tot} = k_i = 0,\, 2\pi/N_s,\,....,\,2\pi
(N_s-1)/N_s$. Then, $\langle i|S_z(q)|j\rangle \neq 0$ only if
$S_z^{\rm tot}(i)=S_z^{\rm tot}(j)$ and $K_{\rm tot}(i) = K_{\rm
tot}(j)+q$. The $T=0$ dynamical structure factor can be obtained
without performing an explicit calculation of all the spectrum of
the Hamiltonian, by using in place of Eq. (1) a continued fraction
representation of the dynamical
susceptibility\cite{Gagliano_Balseiro}, whose coefficients can be
calculated recursively once the ground state is known. We verified
as a check of the calculations that the sum rule $\int d\omega
S^{zz}(q,\omega) = \langle S_z(-q)S_z(q)\rangle$ is verified.\par

Fig.~\ref{Fig_Paolo_spectrum} shows the calculated
$S^{zz}(q,\omega)$ as a function of energy transfer $\omega/J$ for
$q=0.5$ and $q=1$ and for a temperature $T=2J$. The energy spectra
were obtained by convoluting the discrete spectra for finite
chains with Lorentzian functions (width $0.3J$). $q$ is given in
units such that $q=1$ corresponds to the antiferromagnetic point,
which in theoretical work is often referred to as the $\pi$-point.
Throughout the paper, the intensity of the calculated structure
factor $S^{zz}$ is given for a chain of $N_s$ spins and normalized
so that $\sum_{k_i} \int d\omega S^{zz}(k_i,\omega)=
\frac{S(S+1)N_s}{3}$ because in an isotropic system one spin
component carries exactly one third of the intensity.\par

The calculated excitation spectra are broad but still resonant at
$T=2J$, consistent with neutron scattering measurements which
showed that the Haldane excitation survives as a resonant feature
up to at least
$T=2.7J$.\cite{Kenzelmann_CsNiCl3_gap,Kenzelmann_CsNiCl3_temperature}
The calculated scattering weight at $q=1$ is centered at $1.5 J$
so that the scattering is at higher energies than at low
temperature.\cite{Kenzelmann_CsNiCl3_gap} At $q=0.5$, the increase
of the temperature has the opposite effect and the calculated
scattering is centered clearly below the low-temperature maximum
of the dispersion which is $\omega=2.71 J$.\par

The calculated excitation spectra are compared in
Fig.~\ref{Fig_Paolo_spectrum} to neutron scattering data from
${\rm
CsNiCl_3}$.\cite{Kenzelmann_CsNiCl3_gap,Kenzelmann_CsNiCl3_temperature}
The chains in ${\rm CsNiCl_3}$ run along the c-axis, so the third
component $l$ of the wave-vector transfer $\bbox{Q}=(h,k,l)$
corresponds to the wave-vector transfer along the chain. For the
remainder of this paper $q$ denotes both the wave-vector transfer
along a single chain used in the theoretical calculations and the
wave-vector transfer along the chain in ${\rm CsNiCl_3}$, thus
$l=q$. The data in Fig.~\ref{Fig_Paolo_spectrum} was measured at
$T=50\;\mathrm{K}$ which corresponds to $T=2J$.\par

Three-dimensional (3D) spin-spin interactions - due to the
interchain coupling $J'$ - are most important for temperatures
$T<15\;\mathrm{K}$ and they induce antiferromagnetic long-range
order below $T_{\rm N}= 4.84\;\mathrm{K}$.\cite{Morra} Above
$T_{\rm N}$, 3D interactions lead to a dispersion of the
excitations perpendicular to the chain direction,\cite{Morra}
which decreases with increasing temperature but 3D spin
correlations may be still be important at $T=50\;\mathrm{K}$. The
measurements for $q=1$ were performed at a wave-vector transfer
${\bbox Q}_{\rm 1D}=(0.81,0.81,1)$ where the Fourier transform of
the interchain coupling
vanishes\cite{Kenzelmann_CsNiCl3_gap,Kenzelmann_CsNiCl3_temperature}
and where, within a Random Phase Approximation (RPA), the chains
behave as if they were not
coupled.\cite{Buyers86,Nightingale_Blote} So to first order the
experiments probed the dynamics of isolated chains. For $q=0.5$,
the interchain coupling has only a negligible effect on the
dynamic structure factor. Thus it is not important that the
wave-vector transfer perpendicular to the chain varies in energy
scans with the MARI spectrometer and the observed scattering for
$q=0.5$ can be directly compared to the calculations.\par

As shown in Fig.~\ref{Fig_Paolo_spectrum} the calculated and
observed dynamic structure factor are in good agreement for
$q=0.5$ and $q=1$ and the calculated excitation energies and
widths are well predicted by theory. The small differences between
experiment and theory may arise from finite-size effects in the
calculations.\par

\begin{figure}
\begin{center}
  \includegraphics[height=5.5cm,bbllx=60,bblly=265,bburx=490,
  bbury=575,angle=0,clip=]{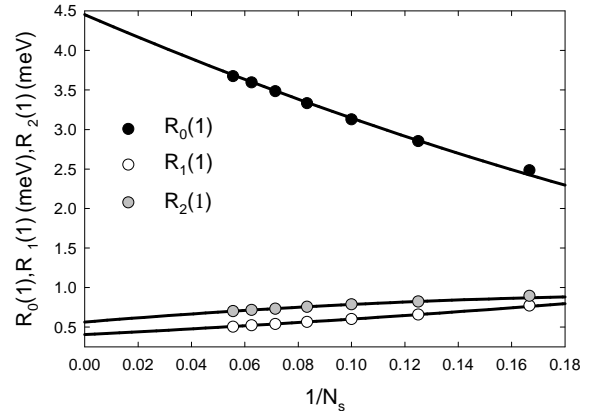}
  \vspace{0.3cm}
  \caption{$R_0(q)=S^{zz}_{\omega>0}(q)$, $R_1(q)=<\omega>_{\omega>0}(q)$
  and $R_2(q)=\sqrt{<\omega^2>_{\omega>0}}(q)$ for $q=1$ obtained
  from ED calculations at zero temperature as a function of the
  inverse chain length $1/N_s$. The solid
  lines are fits to the data to extrapolate to an infinite chain as
  explained in the text.}
  \label{Fig_Paolo_calc}
\end{center}
\end{figure}

The positive frequency moments of the spectrum are expected to be
more reliable than the line-shape of $S^{zz}(q,\omega)$. For
example, at $T=0$ the ratio of the continuum to the Haldane mode
contributions for a single chain is almost independent on the
chain length and is well reproduced in small chain calculations,
despite the fact that the line-shape of the continuum in
$S^{zz}(q=1,\omega)$ is not correct. The following moments were
considered:
\begin{eqnarray}
&R_0(q)=S^{zz}_{\omega>0}(q) = \int_{\omega>0}
S^{zz}(q,\omega)d\omega\, ,&\\\nonumber\\
&R_1(q)=<\omega>_{\omega>0}(q)
=&\nonumber\\&S^{zz}_{\omega>0}(q)^{-1}\int_{\omega>0} \omega
S^{zz}(q,\omega)d\omega\, ,&\\\nonumber\\
&R_2(q)=\sqrt{<\omega^2>_{\omega>0}}(q) =& \nonumber \\
&\left(S^{zz}_{\omega>0}(q)^{-1}\int_{\omega>0} \omega^2
S^{zz}(q,\omega)d\omega\right)^{1/2}\, .&
\end{eqnarray}For $k_{\rm B}T \gtrsim 0.5 J$ finite size
effects in $R_i$ appear to be small
(this is seen by comparing results for $N_s =$ 4,\, 6,\, 8). For
smaller values of $T$, finite-size effects appear to be sizeable
for $q=1$ and very small for $q=0.5$. We extracted the values of
$R_i(q=1)$ at $T=0\;\mathrm{K}$ for $N_s = \infty$ by performing a
second-order finite-size scaling fit:
\begin{equation}
{\tilde R}_i(N_s) = R_i(N_s=\infty) + A_i/N_s + B_i/N_s^2.
\label{Eq-Paolo}
\end{equation}

$A_i$ and $B_i$ were calculated from $R_i$ for $N_s
=14,\,16,\,18$. The extrapolated ${\tilde R}_i(N_s)$ were compared
with the calculated $R_i(N_s)$ for $N_s = 6,\,8,\,10,\,12$ and the
results in Fig.~\ref{Fig_Paolo_calc} show that Eq.~\ref{Eq-Paolo}
holds down to rather small sizes, with just a small deviation for
$R_0(q)(N_s = 6)$ (associated with a small $1/N_s^3$
correction).\par

\section{Quantitative comparison with experiment}

\subsection{Excitations at ${\mathbf q=0.5}$}

\begin{figure}
\begin{center}
  \includegraphics[height=11cm,bbllx=85,bblly=185,bburx=500,
  bbury=722,angle=0,clip=]{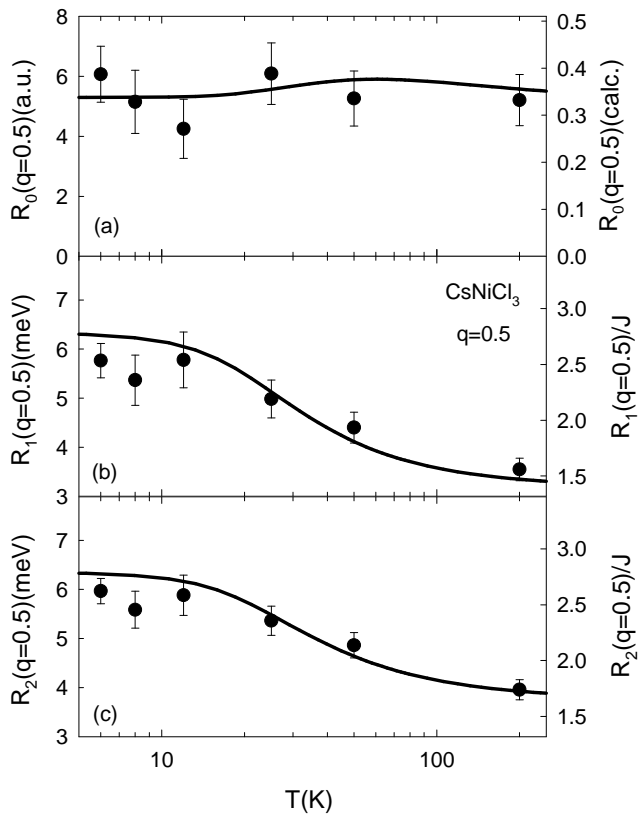}
  \vspace{0.3cm}
  \caption{(a) Integrated neutron scattering intensity $R_0(q)$
  for positive energy transfers at $q=0.5$ as a function of
  temperature on a semi-logarithmic plot. The data were measured
  using MARI and for temperatures between $T=6$ and $200\;\mathrm{K}$
  and taken from Ref.\protect\cite{Kenzelmann_CsNiCl3_temperature}.
  The intensity was determined by adding up the observed spectra for
  positive energy transfers after subtracting the flat background
  and accounting for the magnetic form factor.\protect\cite{Kenzelmann_CsNiCl3_temperature}
  The solid line corresponds to the predicted $R_0(q=0.5)$ obtained
  from ED calculations and fitted to the data, and it is shown in
  units shown of the right axis. (b) First energy moment $R_1(q=0.5)$
  of the neutron scattering spectra observed on the energy loss side.
  (c) Second energy moment $R_2(q=0.5)$ of the scattering spectra on
  the energy loss side. Solid lines in (b-c) are ED predictions.}
  \label{MARI-first-second-halfpi}
\end{center}
\end{figure}

For $q=0.5$, the coupling of the chains in ${\rm CsNiCl_3}$ has a
negligible effect on the dynamic structure factor and a direct
comparison between the experiment and the ED calculations is
possible. Fig.~\ref{MARI-first-second-halfpi} shows the
experimentally observed $R_0(q)$, $R_1(q)$ and $R_2(q)$ for
$q=0.5$ for temperatures between $T=6$ and $200\;\mathrm{K}$. The
data were determined numerically from the neutron spectra measured
using the MARI spectrometer after subtracting the flat background
and accounting for the magnetic form factor of the neutron
scattering intensity.\cite{Kenzelmann_CsNiCl3_temperature}

The ED results for $R_0(q=0.5)$ were adjusted in a least-squares
fit to the experimentally observed intensities using an overall
scaling factor. No scaling factors were used for $R_1(q=0.5)$ and
$R_2(q=0.5)$ which scale with the intrachain exchange
$J=2.28\;\mathrm{meV}$, accurately known from high-field
magnetization measurements.\cite{Katori}\par

The agreement between theory and experiment is excellent as shown
in Fig.~\ref{MARI-first-second-halfpi}. The ED calculations
reproduce the flat temperature dependence of the intensity at
$q=0.5$. A comparison with the overall scale is not possible
because $S({\bbox{Q}},\omega)$ was not measured in absolute units.
Excellent \textit{quantitative} agreement is found for the
temperature dependence of the first and second moment at $q=0.5$,
as shown in Fig.~\ref{MARI-first-second-halfpi}(b)-(c). The small
difference between the experiment and the calculations for
$T<10\;\mathrm{K}$, where $R_1(q=0.5)$ and $R_2(q=0.5)$ are
slightly lower than predicted by our ED calculations, is at least
partly due to finite-size effects: our ED calculations of finite
chains overestimate the energy
$\omega(q=0.5)=2.77J=6.31\;\mathrm{meV}$, whereas a more accurate
zero-temperature calculations give
$\omega(q=0.5)=2.71J=6.17\;\mathrm{meV}$.\cite{Takahashi94}\par
%

\subsection{Excitations at ${\mathbf q=1}$}
\begin{figure}
\begin{center}
  \includegraphics[height=11cm,bbllx=63,bblly=131,bburx=470,
  bbury=653,angle=0,clip=]{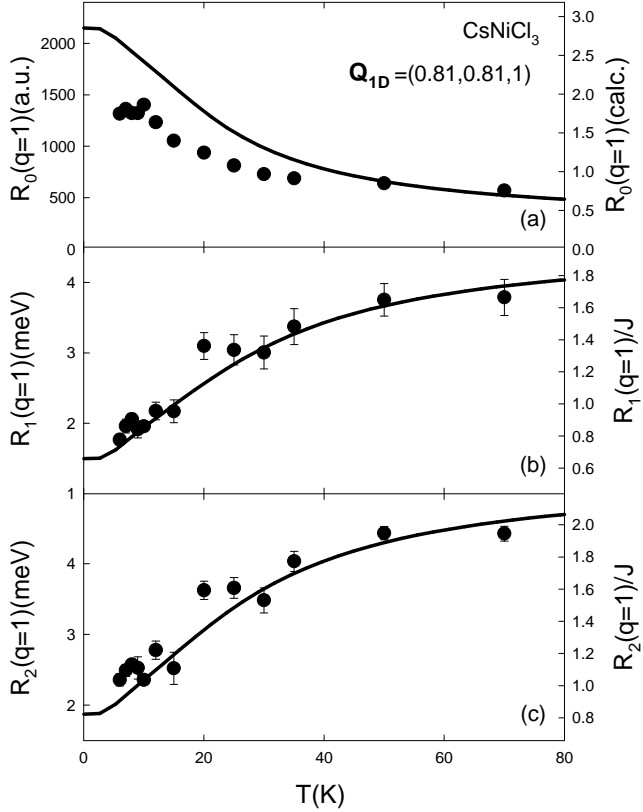}
  \vspace{0.3cm}
  \caption{(a) Integrated neutron scattering intensity
  $R_0(q=1)$ observed at the non-interacting wave-vector
  ${\bbox Q}_{\rm 1D}=(0.81, 0.81, 1)$ using the RITA spectrometer
  and taken from Ref.\protect\cite{Kenzelmann_CsNiCl3_temperature}.
  The solid line corresponds to the calculated $R_0(q=1)$ obtained
  by ED calculations. The prediction was scaled to match the
  intensity at $T=70\;\mathrm{K}$ and is shown in units shown on
  the right axis. The error bars are smaller than the symbol size.
  (b-c) First and second energy moment $R_1(q=1)$ and
  $R_2(q=1)$ for the energy loss neutron scattering
  spectrum observed at ${\bbox Q}_{\rm 1D}$ using the RITA
  spectrometer. The solid line was obtained by ED
  calculations and the temperature and energy scaling was
  done using the intra-chain exchange $J=2.28\;\mathrm{meV}$
  known from high-field magnetization measurements.}
  \label{RITA-first-second}
\end{center}
\end{figure}

Fig.~\ref{RITA-first-second} shows $R_0(q)$, $R_1(q)$ and $R_2(q)$
for $q=1$, measured between $T=6$ and $70\;\mathrm{K}$ using the
RITA spectrometer and compared to ED calculations of a finite
chain of $8$ sites. The neutron scattering measurements were
performed at the non-interacting wave-vector ${\bbox Q}_{\rm
1D}$\cite{Kenzelmann_CsNiCl3_gap,Kenzelmann_CsNiCl3_temperature}
where the chains behave within RPA as if they were not coupled. To
first order, the temperature dependence of $R_0(q=1)$, $R_1(q=1)$
and $R_2(q=1)$ are thus expected to match that of a single
chain.\par

$R_1(q=1)$ is about $0.65J$ at zero temperature, and it is almost
three times as large at $T=2J$. This reflects the rapid increase
of the excitation energy at $q=1$ with increasing
temperature.\cite{Kenzelmann_CsNiCl3_gap} The zero-temperature
value of $R_1(q=1)$ is higher than the Haldane gap $0.41J$ due to
the presence of a weak single-chain multi-particle continuum at
higher energies.\cite{Horton_Affleck,Essler} There is good
agreement between the measured and calculated $R_1(q=1)$ and
$R_2(q=1)$ for temperatures $T>15\;\mathrm{K}$ as shown in
Fig.~\ref{RITA-first-second}. Below this temperature, the
experimentally observed $R_1(q=1)$ and $R_2(q=1)$ are consistently
higher then predicted by the calculations. In this temperature
regime, finite-size effects are important which make the
calculated $R_1(q=1)$ and $R_2(q=1)$ (see
Fig.~\ref{Fig_Paolo_calc}) larger than those of an infinite chain
and finite-size effects can thus not explain the difference
between the experiment and the calculations.\par

The difference between the experiment and the ED calculations are
probably due to interchain couplings, which are known to have a
strong effect on the intensity of the excitations just above the
ordering temperature, as shown in Fig. 5 of our experimental
paper.\cite{Kenzelmann_CsNiCl3_temperature} Also, for temperatures
$T<12\;\mathrm{K}$ a multi-particle continuum was
observed,\cite{Kenzelmann_CsNiCl3_continuum,Kenzelmann_CsNiCl3_continuum_long}
which increases $R_1(q=1)$ and $R_2(q=1)$ due to the presence of
scattering weight at high energy transfers.\par

\begin{figure}
\begin{center}
  \includegraphics[height=5.8cm,bbllx=78,bblly=257,bburx=540,
  bbury=569,angle=0,clip=]{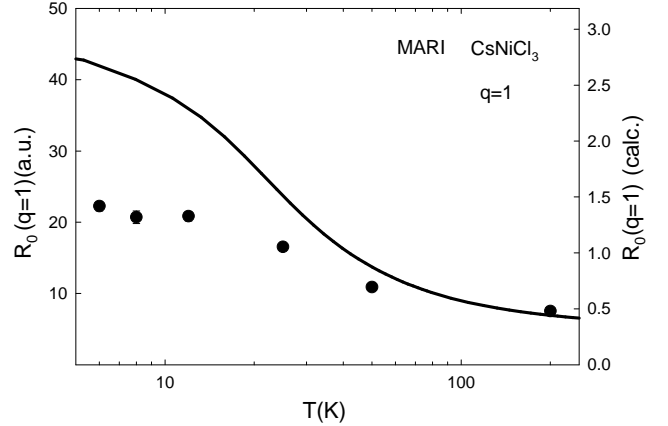}
  \vspace{0.3cm}
  \caption{Neutron scattering intensity at $q=1$
  observed using MARI for temperatures between $T=6$ and
  $200\;\mathrm{K}$ as solid circles on a semi-logarithmic
  plot and taken from Ref.\protect\cite{Kenzelmann_CsNiCl3_temperature}.
  The intensity was determined by integrating the
  observed spectra for positive energy transfers after
  subtracting a flat background and correcting for the
  magnetic form factor.\protect\cite{Kenzelmann_CsNiCl3_temperature}
  The solid line is the ED result for $R_0(q)$ at $q=1$. The
  scaling was chosen to match the prediction $R_0(q=1)/R_0(q=0.5)$
  and is thus in the same units as the scaling in
  Fig.~\ref{MARI-first-second-halfpi}.}
  \label{MARI-intensity-pi}
\end{center}
\end{figure}

Figs.~\ref{RITA-first-second}(a) and \ref{MARI-intensity-pi} show
the experimentally observed $R_0(q)$ at $q=1$ as a function of
temperature. The data were measured using the RITA triple-axis and
the MARI time-of-flight spectrometer,
respectively.\cite{Kenzelmann_CsNiCl3_gap,Kenzelmann_CsNiCl3_temperature}
The ED calculations predict that at $T=7.5J$ the ratio
$R_0(q=1)/R_0(q=0.5)$ is equal to $1.26$ and this can directly be
tested with the neutron scattering experiments. At $T=7.5J \simeq
200\;\mathrm{K}$ the MARI measurements show that
$R_0(q=1)/R_0(q=0.5)=1.5(4)$, in good agreement with the
calculations. At lower temperatures, however, the experimentally
observed $R_0(q=1)/R_0(q=0.5)$ is less than predicted and this is
shown in Fig.~\ref{MARI-intensity-pi}. Since the temperature
dependence of $R_0(q=0.5)$ is in agreement with the calculations
(Fig.~\ref{MARI-first-second-halfpi}), we conclude that the
disagreement results from a reduced intensity $R_0(q)$ at the
antiferromagnetic point $q=1$. The disagreement is most pronounced
for $T<50\;\mathrm{K}$ and may be due to the fact that the
measurements were performed close to but not exactly at the
non-interacting wave-vector ${\bbox Q}_{\rm 1D}$.\par

Similar result were obtained, however, from an analysis of the
RITA measurements which were performed \textit{exactly} at the
non-interacting 1D point ${\bbox Q}_{\rm 1D}$ in reciprocal space
so that to first order the neutron spectrum should not be
influenced by 3D interactions in the system. For the RITA
measurements the scaling factor for $R_0(q=1)$ can only be guessed
because reliable measurements at $q=0.5$ were not performed.
Choosing the scaling factor such that the ED results match the
observed $R_0(q=1)$ at $T=70\;\mathrm{K}$ - which is known to be
approximately true from the MARI measurements - the solid line
shown in Fig.~\ref{RITA-first-second}(a) is obtained. The
discrepancy between the experiment and the numerical calculations
is similar to that observed using the MARI spectrometer. At low
temperatures there is a difference between theory and experiment
of a factor of $1.6$.\par

Finite-size effects may be important for $T<15\;\mathrm{K}$, but
these would underestimate the measured intensity rather than
overestimate it (Fig.~\ref{Fig_Paolo_calc}) and so they cannot
explain the discrepancy between the experiment and the numerical
calculations. This means the difference between the experiment
would be even bigger if longer chains had been considered in the
calculations and the observed $R_0(q=1)$ is \textit{at least} a
factor of $1.6$ less than that calculated.\par

\subsection{The influence of next-nearest neighbor, biquadratic
and interchain spin-spin interactions}

We examined the effect of next-nearest neighbor $J_{\rm nn}$ and
biquadratic $J_{\rm biq}$ exchange interactions on the dynamic
structure factor at zero temperature. The maximum in the
dispersion $\omega(q=0.5)$ was calculated numerically for a spin-1
Heisenberg chain with Hamiltonian ${\mathcal{H}}$
(Eq.~\ref{Hamiltonian1D-diag}) and with additional terms
\begin{equation}
     {\mathcal H}_{\rm nn} = J_{\rm nn}\sum_{i} \bbox{S}_{\rm i} \cdot
     \bbox{S}_{\rm i+2}
\end{equation}for next-nearest neighbor interactions or
\begin{equation}
     {\mathcal H}_{\rm biq} = J_{\rm biq} \sum_{i} (\bbox{S}_{\rm i} \cdot
     \bbox{S}_{\rm i+1})^2
\end{equation}for biquadratic spin interactions. The calculations were
performed for chains with $12$ sites. The calculated energies were
then compared with the experimentally observed energies for
$q=0.5$ at low temperatures in order to estimate the strengths of
next-nearest neighbor and biquadratic exchange interactions.\par

\begin{figure}
\begin{center}
  \includegraphics[height=10.7cm,bbllx=104,bblly=142,bburx=500,
  bbury=700,angle=0,clip=]{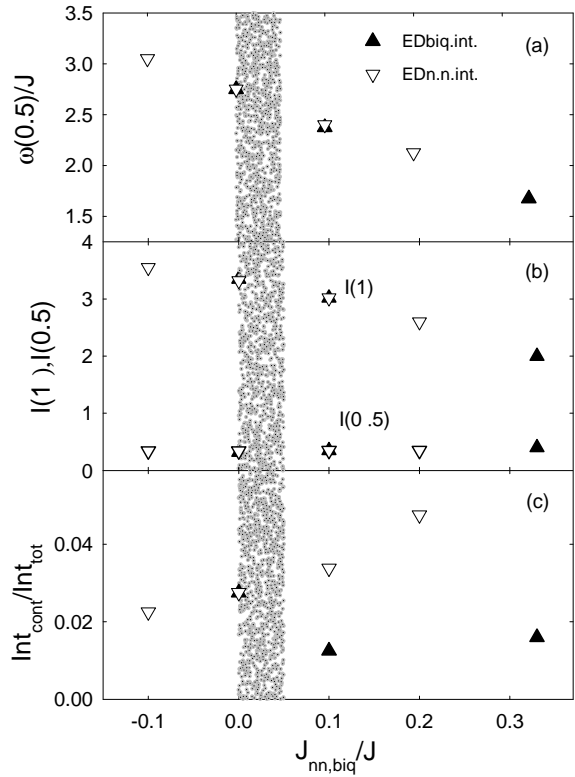}
  \vspace{0.3cm}
  \caption{(a) Zero-temperature ED results for the excitation
  energy at $q=0.5$ as function of $J_{\rm nn}/J$ and
  $J_{\rm biq}/J$. (b) ED results for the integrated intensity
  at $q=1$ and at $q=0.5$ at zero temperature as a
  function of $J_{\rm nn}/J$ and $J_{\rm biq}/J$. (c)
  Zero-temperature results for the ratio of continuum
  intensity $I_{\rm cont}$ to total intensity $I_{\rm tot}$ at
  $q=1$ as a function of $J_{\rm nn}/J$ and $J_{\rm biq}/J$.
  The shaded area in (a)-(c) correspond to the possible
  $J_{\rm nn}/J$ and $J_{\rm biq}/J$ range for ${\rm CsNiCl_3}$.}
  \label{paolo-nnn-biq}
\end{center}
\end{figure}

$\omega(q=0.5)$ decreases for increasing $J_{\rm nn}/J$ and
$J_{\rm biq}/J$ and increases when these exchange couplings are
negative, as shown in Fig.~\ref{paolo-nnn-biq}(a). The measured
values for $\omega(q=0.5)$ are between $5.9$ and
$6.2\;\mathrm{meV}$ for various experiments with
$6\;\mathrm{K}<T<12\;\mathrm{K}$
(Ref.~\cite{Kenzelmann_CsNiCl3_temperature,Morra,Kenzelmann_CsNiCl3_continuum_long}).
This is consistent with $0<J_{\rm nn}/J,J_{\rm biq}/J<0.05$ and
this range is shown in Fig.~\ref{paolo-nnn-biq}(a)-(c) as shaded
area. The associated decrease in $R_0(q=1)$ is at most a factor of
$1.05$ (Fig.~\ref{paolo-nnn-biq}(b)), which is much smaller than
the observed reduction of $R_0(q=1)$ by a factor of $1.6$ at low
temperatures. This shows that next-nearest neighbor and
biquadratic exchange interactions cannot lead to the
experimentally observed decrease in $R_0(q=1)$.\par

Fig.~\ref{paolo-nnn-biq}(c) shows that the recently observed
multi-particle continuum\cite{Kenzelmann_CsNiCl3_continuum} at
high energy transfer carrying $12\%$ of the total scattering at
$q=1$ cannot be caused by biquadratic or next-nearest neighbor
interactions. This is because for $0<J_{\rm nn}/J,J_{\rm
biq}/J<0.05$ the strength of the continuum is never more than
$3\%$ of the total scattering at $q=1$. We note that our results
for the valence bond solid model ($J_{\rm biq}/J=0.33$) are
consistent with a previous study.\cite{Takahashi94}\par

At low temperatures $T<12\;\mathrm{K}$, the observed reduction of
$R_0(q=1)$ for ${\bbox Q}_{\rm 1D}$ is related to the correlation
length $\xi$ of the coherent state in the chains. It was shown
that in ${\rm CsNiCl_3}$ $\xi$ is reduced from the value expected
for single chains,\cite{Kenzelmann_CsNiCl3_temperature} which
results in a wider peak of the structure factor $S(q)$ centered at
$q=1$. This means that $S({\bbox Q}_{\rm 1D})$ is reduced with
respect to the value for uncoupled chains if the $q$-integrated
intensity of $S(q)$ is to be unchanged.\par

The reduction of $R_0(q=1)$ is also related to the multi-particle
continuum at higher
energies:\cite{Kenzelmann_CsNiCl3_continuum,Kenzelmann_CsNiCl3_continuum_long}
The Hohenberg-Brinkman sum rule predicts that the first energy
moment $F(q)=\int d\omega\, \omega\, S(q,\omega)$ of
one-dimensional magnets with nearest-neighbor exchange interaction
behaves as\cite{Hohenberg_Brinkman}
\begin{equation}
     F(q)= -\frac{4}{3} <{\mathcal{H}}> (1-\cos(\pi q))\, .
     \label{Eq-HB}
\end{equation}It will be shown in the Appendix that this relation
also holds for coupled chains with small interchain interactions.
If the integrated intensity for a particular $q$ is reduced,
intensity necessarily has to be transferred to high energies so
that the spectrum is consistent with the Hohenberg-Brinkman sum
rule. This is because intensity at high energies contributes more
to the first moment than scattering at low energies.\par

The comparison between the experiment and numerical calculations
suggests that 3D correlations lower the intensity at the
non-interacting 1D point ${\bbox Q}_{\rm 1D}$ for $T<2J$.
Conservation of the total magnetic intensity then implies that
this intensity is transferred to other parts of reciprocal space.
The 3D dependence of the magnetic scattering suggests a coupling
of domain walls on different chains for temperatures up to at
least $T=2J$. Further the coupling of the chains in
antiferromagnetic spin-1 Heisenberg chains has a larger effect on
the excitation spectrum and extends to higher temperatures than
assumed so far. Possibly the hidden string order plays a decisive
part in this effect because it allows the existence of
spatially-extended coherent states to higher temperatures than
would be possible in magnets without a similar spin string
order.\par

\section{Temperature dependence of \mbox{\boldmath $<{\mathcal{H}}>$}}

In our experimental paper\cite{Kenzelmann_CsNiCl3_temperature} the
first energy moment sum rule was used to determine the temperature
dependence of the expectation value of the Hamiltonian
$<{\mathcal{H}}>$ of a single spin-1 chain. Here we show that the
first moment sum rule for single chains is also valid for weakly
coupled chains. We also compare the experimentally determined
values for $<{\mathcal{H}}>$ to ED results.\par

The first energy moment of ${\rm CsNiCl_3}$ based on the
Hamiltonian given in Eq.~\ref{Hamiltonian-diag} can be written as
(see the Appendix)
\begin{eqnarray}\label{Eq_1stmom_3D}
&F(\bbox{Q})= - 4 J F (1 - \cos(\bbox{Q} \cdot
\bbox{c}/2))&\nonumber
\\ \nonumber
\\ & - 4 J' F' {\displaystyle \sum_j} (1-\cos(\bbox{Q} \cdot \bbox{a}_j))\, ,&
\end{eqnarray}where $a_j$ are the lattice positions of the nearest
neighbors and
\begin{eqnarray}
&F=F_{\alpha} = {\displaystyle \sum_i}<S_{\bbox{r}_i}^{\alpha}
S_{\bbox{r}_i +\bbox{c}/2}^{\alpha}>,&\nonumber \\ \nonumber \\
&F'=F'_{\alpha} = {\displaystyle \sum_i}<S_{\bbox{r}_i}^{\alpha}
S_{\bbox{r}_i +\bbox{a}_j}^{\alpha}>,&\nonumber
\end{eqnarray}for $\alpha=x,\,y,\,z$. The second term in
Eq.~\ref{Eq_1stmom_3D} is always much smaller than the first term,
mainly because $J'=0.015J$, but also because ${\rm Max}\{\sum_j
(1-\cos(\bbox{Q} \cdot \bbox{a}_j))\}=4.5$ is only 2.25 times
${\rm Max}\{(1 - \cos(\bbox{Q} \cdot \bbox{c}/2))\}=2$. In
addition, the wave-vector transfer in the experiments was mainly
along the $c$-axis, which reduces the second term.\par

The dominant term in Eq.~\ref{Eq_1stmom_3D} is therefore
\begin{eqnarray}
&F(\bbox{Q})= - 4 J F (1 - \cos(\bbox{Q} \cdot
\bbox{c}/2))&\nonumber
\\ \nonumber \\ &  = - 4 J F (1 - \cos(\pi q))\, ,&
\label{Eq_first_moment}
\end{eqnarray}which is directly proportional to $<{\mathcal
H}>=3JF$ with ${\mathcal H}$ given in
Eq.~\ref{Hamiltonian1D-diag}. The expectation value of the
Hamiltonian for a single chain can thus be obtained by fitting
Eq.~\ref{Eq_first_moment} to the $q$-dependence of the
experimentally observed first energy moment
$F(q)$.\cite{Kenzelmann_CsNiCl3_temperature}\par

\begin{figure}
\begin{center}
  \includegraphics[height=5.5cm,bbllx=80,bblly=290,bburx=540,
  bbury=603,angle=0,clip=]{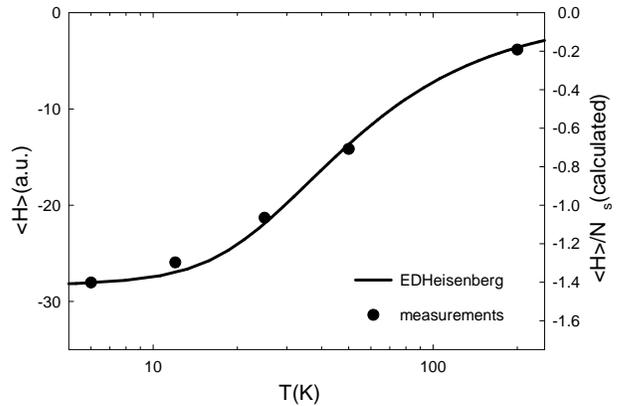}
  \vspace{0.3cm}
  \caption{The experimentally observed $<{\mathcal{H}}>$ =
  $J \sum_i <S_i S_{i+1}>$ as solid circles and as a function
  of temperature on a semi-logarithmic plot and taken from
  Ref.\protect\cite{Kenzelmann_CsNiCl3_temperature}.
  $<{\mathcal{H}}>$ was determined from the first energy
  moment measured at the different temperatures. The error
  bars are smaller than the symbol size. The solid line is
  the ED result for the temperature dependence of the mean
  energy per spin $<{\mathcal{H}}>/N_s$.}
  \label{Fig-gs-energy}
\end{center}
\end{figure}

Fig.~\ref{Fig-gs-energy} shows the experimentally determined
$<{\mathcal{H}}>$ from Ref.~\cite{Kenzelmann_CsNiCl3_temperature},
compared with our ED calculations. We find that there is excellent
agreement between theory and experiment for the whole temperature
range. A scaling factor was used to reach the agreement since
$<{\mathcal{H}}>$ was not measured in absolute units.\par

The calculated ground state energy per spin at zero temperature,
$<{\mathcal H}>/N_s=-1.41(1) J$, is consistent with a previous
numerical study of much longer chains.\cite{White_Huse}\par

\section{Conclusions}
We have performed exact diagonalization calculations of a finite
antiferromagnetic spin-1 chain with $8$ sites. The calculations
confirm the recent experimental observation that the excitations
are resonant for temperatures up to at least $T=2J$. The
temperature dependence of the calculated first and second energy
moments for positive energy transfers is in good agreement with
the measured neutron
scattering\cite{Kenzelmann_CsNiCl3_temperature} except at low
temperatures where spin correlations between chains are most
important. Here the measured first and second moment for $q=1$ are
higher than predicted due to the presence of a multi-particle
continuum at high energy transfers.\par

The measured intensity of the excitations at the non-interacting
wave-vector is much weaker than that calculated for isolated
chains, particularly for $T<2J$. This suggests that solitons on
different chains interact for temperatures up to at least $T=2J$.
The experimentally observed $<{\mathcal H}>$ is in excellent
agreement with exact diagonalization calculations for the entire
temperature range considered in this study ($0.25J < T
<7.5J$).\par

\section*{Appendix}
The first energy moment is defined as
\begin{eqnarray}
&F^{\alpha\alpha}(\bbox{Q}) =
\int_{\infty}^{\infty}\frac{d\omega}{2\pi} \omega S^{\alpha
\alpha}(\bbox{Q},\omega) =&\nonumber \\ \nonumber \\&
-\frac{1}{2\hbar} <\left[\left[{\mathcal{H}},
S^\alpha_{\bbox{Q}}\right] ,S^\alpha_{-\bbox{Q}}\right]>\, ,&
\end{eqnarray}where the Fourier transform of the spin operators
$S_{\bbox{r}_i}^{\alpha}$ is defined as
\begin{equation}
S^\alpha_{\bbox{Q}}=N_s^{-1/2}\sum_i
\exp(-i\bbox{Q}\bbox{r}_i)S_{\bbox{r}_i}^{\alpha}\, .\nonumber
\end{equation}In the the case of ${\rm CsNiCl_3}$, the
Hamiltonian includes both intrachain and interchain spin
interaction, and it can be written as
\begin{eqnarray}
&{\mathcal{H}}={\displaystyle \sum_{i,j}} J_x S^x_{\bbox{r}_i}
S^x_{\bbox{r}_i+\bbox{c}/2} & \nonumber\\ \nonumber \\&  + J_y
S^y_{\bbox{r}_i} S^y_{\bbox{r}_i+\bbox{c}/2} + J_z
S^z_{\bbox{r}_i} S^z_{\bbox{r}_i+\bbox{c}/2}&\nonumber \\
\nonumber \\ &+ J'_x S^x_{\bbox{r}_i} S^x_{\bbox{r}_i+\bbox{a}_j}
+ J'_y S^y_{\bbox{r}_i} S^y_{\bbox{r}_i+\bbox{a}_j} + J'_z
S^z_{\bbox{r}_i} S^z_{\bbox{r}_i+\bbox{a}_j}\, .&\nonumber \\
\end{eqnarray}
In a cartesian coordinate system, the positions of the neighbors
along the three directions in the basal plane are given as
\begin{eqnarray}
&\bbox{a}_1=(a,0,0),&\nonumber\\
&\bbox{a}_2=(a/2,\sqrt{3}a/2,0),&\nonumber
\\ &\bbox{a}_3=(-a/2,\sqrt{3}a/2,0)\, .&\nonumber
\end{eqnarray}The first moment for $S^{xx}(\bbox{Q},\omega)$
can be split up into two terms, one containing the intrachain
interactions $J_\alpha$ and the other the inter-chain interaction
$J'_\alpha$:
\begin{eqnarray}
&F^{xx}(\bbox{Q})= - 2J_y (F_y - F_z \cos(\bbox{Q} \cdot
\bbox{c}/2))&\nonumber \\ \nonumber \\ &- 2J_z (F_z - F_y
\cos(\bbox{Q} \cdot \bbox{c}/2))&\nonumber\\ \nonumber \\& -
{\displaystyle \sum_j} \left[2J'_y (F'_{y,j} - F'_{z,j}
\cos(\bbox{Q} \cdot \bbox{a}_j))\right.&\nonumber \\ \nonumber \\
&\left. + 2J'_z (F'_{z,j} - F'_{y,j} \cos(\bbox{Q} \cdot
\bbox{a}_j))\right]\, .&
\end{eqnarray}$F_{\alpha}$ and $F'_{\alpha,j}$ are the expectation
values for spin-spin correlations along the chain direction and
along either of the three direction in the basal plane:
\begin{eqnarray}
&F_{\alpha} = {\displaystyle \sum_i} <S_{\bbox{r}_i}^{\alpha}
S_{\bbox{r}_i +\bbox{c}/2}^{\alpha}>,&\nonumber\\ \nonumber
\\&F'_{\alpha,j} = {\displaystyle \sum_i} <S_{\bbox{r}_i}^{\alpha}
S_{\bbox{r}_i +\bbox{a}_j}^{\alpha}>\, .&\nonumber
\end{eqnarray}Due to the discrete rotational symmetry $D^4_{6h}$
in the magnetically disordered phase, spin correlations along all
three equivalent direction in the basal plane are the same and
\begin{eqnarray}
&F'_{\alpha} = F'_{\alpha,j},&\hspace{1cm} j=1,2,3\, .\nonumber
\end{eqnarray}For an isotropic spin chain $J_{\alpha}=J$, $F_{\alpha}=F$
and $F_{\alpha}'=F'$ for $\alpha=x,\,y,\,z$ and the first moment
of $S(\bbox{Q},\omega)$ is
\begin{eqnarray}
&F(\bbox{Q})= - 4 J F (1 - \cos(\bbox{Q} \cdot
\bbox{c}/2))&\nonumber
\\ \nonumber \\ & - 4 J' F' {\displaystyle \sum_j} (1-\cos(\bbox{Q} \cdot \bbox{a}_j))\, .&
\end{eqnarray}

\begin{acknowledgments}
We would like to thank Prof. R.~A. Cowley and Dr. W.~J.~L. Buyers
for enlightening discussions. The authors were supported by the
Swiss National Science Foundation under Contract No. 83EU-053223
and No. 8220-053437, respectively.
\end{acknowledgments}

\bibliographystyle{prsty}

\end{document}